\documentclass[aps,showpacs,preprintnumbers,amsmath,amssymb,nofootinbib,eqsecnum,twocolumn,preprintnumbers]{revtex4}
\usepackage{graphicx}
\usepackage{epsf}
\usepackage{amsmath}
\usepackage{epstopdf}
\usepackage{bm}
\usepackage{color}
\usepackage{tabularx}
\usepackage{enumitem}
\usepackage{float}
\usepackage{array,booktabs}
\usepackage{footnote}
\usepackage{threeparttable}
\usepackage{graphicx}
\usepackage{hyperref}
\usepackage{amssymb,epsf}
\usepackage{latexsym}
\usepackage{epstopdf}
\usepackage{epsfig}
\usepackage{eurosym}
\usepackage{amsfonts}
\usepackage{amssymb}
\usepackage{xcolor}
\usepackage{subfigure}

\begin{document}
 \title{Quantum Cosmology in Thick Brane}
 \author{
  S. H. Hendi$^{1,2,4}$ \footnote{email address: hendi@shirazu.ac.ir}
 N. Riazi$^{3}$ \footnote{email address: n$\_$riazi@sbu.ac.ir} and
 S. N. Sajadi$^{1,2}$\footnote{email address: naseh.sajadi@gmail.com},
 }
 \affiliation{
 $^1$Department of Physics, School of Science, Shiraz University, Shiraz 71454, Iran \\
 $^2$Biruni Observatory, School of Science, Shiraz University, Shiraz 71454, Iran    \\
 $^3$Department of Physics, Shahid Beheshti University, G.C., Evin, Tehran 19839, Iran    \\
 $^4$Canadian Quantum Research Center 204-3002 32 Ave Vernon, BC V1T 2L7 Canada
 }

\begin{abstract}
One of the interesting open problems in the cosmological framework
is applying quantum physics to the whole universe, consistently.
Although different conceptual aspects of quantum cosmology are
still very much alive, till now, all the attempts of obtaining a
unique and well-defined version of quantum cosmology have been
unfeasible. Motivated by what was mentioned above, in this paper,
the quantum tunneling of thick brane is investigated through
deriving the Wheeler-DeWitt equation and corresponding solutions.
Besides, the cosmological analysis of the brane is studied based
on two special cases of the scalar field. Finally, it is found
that the constant scalar potential arising from a time-dependent
scalar field is supported by the so-called slow-roll inflation.

\end{abstract}

\maketitle

\section{Introduction}

One of the cornerstones of theoretical physics is the
determination of spacetime dimension. In other words, it is an
open question that whether the spacetime dimension is really $3+1$
or it might be, fundamentally, higher than four. The
Kaluza---Klein (KK) theory is the first serious attempt beyond the
usual four-dimensional spacetime. The fifteen components of the
five-dimensional KK metric are divided into ten components
identifying the four-dimensional spacetime metric, a $U(1)$ gauge
field with four components, and one scalar field called the
dilaton. Although it is argued that the approach of KK theory is
not accurate, the concept of extra dimension in such a theory
provided a basis for further hot topics such as superstring
theory, brane cosmology, and AdS/CFT correspondence or its
generalization to gauge/gravity duality.

The key point of the braneworld scenario is the fact that the
$(3+1)$-dimensional brane is embedded in a five-dimensional
geometry in which the extra dimension may be large or compact.
Regarding the effective models of brane cosmology such as the
Randall and Sundrum \cite{RS} involving an extra-dimension with
non-trivial warp factor, it is possible to discuss the hierarchy
problem to explain the weakness of gravity relative to the other
forces of nature. Indeed, it is supposed that our visible universe
is a membrane inside a higher-dimensional space, and particles and
interactions corresponding to electromagnetic, weak, and strong
forces are restricted to such a membrane.

In this paper, we take into account a five-dimensional space.
There are two known cases in this direction. As the first case,
one may consider an infinitely thin brane-universe along the
extra-dimension. The validity of this option depends on the energy
scales that one wishes to examine. The second case is devoted to
the brane with finite thickness. Hereafter, we consider a thick
brane model by assuming that the energy-momentum tensor describing
the brane has a finite distribution over a limited length interval
along the fifth dimension.

The first approach to describe the universe in the context of
quantum theory was reported in $1960$ by Wheeler \cite{13} and
DeWitt \cite{14}. They proposed a quantum gravity equation to
describe the wave function of the universe, which is known as the
Wheeler-DeWitt (WD) equation. This equation is analogous to a
zero-energy Schr\"{o}dinger equation in which the Hamiltonian
contains both the gravitation and scalar fields. However, one of
the most important problems for solving the WD equation is the
subject of initial conditions. Unlike a classical system, for
cosmological models, there are no external initial conditions
because there is no external time parameter to the universe. To
solve this problem, two different approaches are known: the
Hartle-Hawking no-boundary \cite{15}-\cite{18} and the Vilenkin
tunneling proposal \cite{19}-\cite{22}. The first proposal is
based on an assumption that the wave function of the universe is
given by a path integral over compact Euclidean geometries, and
therefore, this universe has no boundary in this space. The second
one states that the universe spontaneously nucleates and then
evolves along the lines of an inflationary scenario. The
mathematical description of this approach is closely analogous to
that of quantum tunneling through a potential barrier. Indeed,
only the outgoing modes of the wave function should be taken at
the singular boundary of superspace \cite{22,23}.

Working in the framework of brane's theory, the main motivation of
this paper is to understand the earliest moments of the universe,
where we expect that quantum effects are dominant. This paper is
organized as follows: In the next section, the canonical
quantization is applied on the thick brane, the WD equation is
obtained and its solution is investigated. Section \ref{Cos} is
devoted to the cosmology of thick brane and finally in Sec.
\ref{Con} concluding remarks are presented.

\section{Canonical Quantum Cosmology\label{FE}}

As the starting point, we consider the usual five-dimensional
Einstein-scalar field theory with the following bulk action
\begin{equation}
S=\int d^{5}x\sqrt{g^{(5)}}\left( R^{(5)}-\dfrac{1}{2}\partial
_{A}\phi
\partial ^{A}\phi -V(\phi )\right),  \label{ac}
\end{equation}
where the bulk indices $A$ and $B$ runs through $0,1,2,3,4$ i.e.
over all five-dimensional spacetime. In order to obtain the field
equations, one may use the variation of the action with respect to
the bulk metric $g_{AB}$ and scalar field $\phi$ to achieve
\begin{eqnarray}
R_{AB}-\dfrac{1}{2}g_{AB}R^{(5)} &=&T^{\phi}_{AB},  \label{eqmotion1} \\
\square _{5}\phi &=&\dfrac{dV}{d\phi },  \label{eqmotion2}
\end{eqnarray}
where $T^{\phi}_{AB}=\partial _{A}\phi \partial _{B}\phi
-g_{AB}\left( \dfrac{1}{2}\partial _{C}\phi \partial ^{C}\phi
+V(\phi )\right)$ and $\square _{5}\phi =g^{AB}\phi _{;AB}$. At
this point, we have to consider the geometrical properties of the
spacetime by introducing a metric. We adopt the following metric
ansatz to describe the five-dimensional spacetime \cite{Hendi:2020qkk}
\begin{widetext}
\begin{equation}
ds^{2}=a^{2}(w)\left[ -dt^{2}+b^{2}(t)\left( \dfrac{dr^{2}}{1-\kappa r^{2}}%
+r^{2}\left[ d\theta ^{2}+sin^{2}(\theta )d\phi ^{2}\right]
\right) \right] +dw^{2},  \label{eqmetric}
\end{equation}
\end{widetext}
where $a(w)$ is the warp function of the brane, $b(t)$ is the
brane scale factor and $\kappa $ is related to the scalar
curvature. Now, we wish to quantize the system described by the
action (\ref{ac}) through the canonical quantization. To
investigate the canonical quantization, one should promote the
metric $g_{ij}$, the conjugate momenta $\Pi _{ij}$, the
Hamiltonian density $H$ and the momentum density $H_{i}$ to
quantum operators satisfying canonical commutation relations.

At first, we should calculate the Lagrangian term by term. The
Ricci scalar
of the mentioned metric is given by%
\begin{equation}
R^{(5)}=-\dfrac{2\left[ 4ab^{2}a^{^{\prime \prime }}+6b^{2}a^{^{\prime }2}-3b%
\ddot{b}-3\dot{b}^{2}-3\kappa \right] }{a^{2}b^{2}},  \label{R5}
\end{equation}%
where $a=a(w)$, $b=b(t)$. Also dot and prim are derivatives with respect to $%
t$ and $w$, respectively. By using Eq. (\ref{ac}), the Lagrangian
is
\begin{multline}
L=-2a^{2}b^{3}\left[ \frac{3\dot{b}^{2}}{b^{2}}+6a^{^{\prime
}2}+4aa^{^{\prime \prime
}}-\dfrac{1}{4}\dot{\phi}(w,t)^{2}+\right.  \notag
\\
\left. \dfrac{1}{4}a^{2}\phi ^{^{\prime }}(w,t)^{2}+\frac{1}{2}a^{2}V-\frac{%
3\kappa }{b^{2}}\right] .\hspace{6cm}  \label{ll}
\end{multline}%
\newline
In addition, the momenta conjugate to $b$ and $\phi $ are%
\begin{eqnarray}
\Pi _{b} &=&\dfrac{\partial L}{\partial \dot{b}}=-12a^{2}b\dot{b},
\label{Pi-b} \\
\Pi _{\phi } &=&\dfrac{\partial L}{\partial
\dot{\phi}}=a^{2}b^{3}\dot{\phi}. \label{Pi-phi}
\end{eqnarray}%
Now, we can calculate the Hamiltonian with the following explicit form%
\begin{equation}
H=\Pi _{b}\dot{b}+\Pi _{\phi }\dot{\phi}-L=-\Pi _{b}^{2}+\dfrac{12}{b^{2}}%
\Pi _{\phi }^{2}+U(a,b,\phi )=0,  \label{H}
\end{equation}%
where $U(a,b,\phi )$ is%
\begin{equation}
U(a,b,\phi )=24a^{4}b^{4}\left( 12a^{^{\prime }2}+8aa^{^{\prime \prime }}+%
\dfrac{a^{2}\phi ^{^{\prime }2}}{2}+a^{2}V-\frac{6\kappa
}{b^{2}}\right) . \label{eqsuperpot}
\end{equation}

By taking the replacement $\Pi _{b}\rightarrow -i\dfrac{\partial }{\partial b%
}$ and $\Pi _{\phi }\rightarrow -i\dfrac{\partial }{\partial \phi
}$ and
imposing $H\Psi =0$, one can find the following WD equation%
\begin{equation}
\left( \dfrac{\partial ^{2}}{\partial b^{2}}-\dfrac{12}{b^{2}}\dfrac{%
\partial ^{2}}{\partial \phi ^{2}}+U(a,b,\phi )\right) \Psi (b,\phi )=0,
\label{WD-eq}
\end{equation}%
which has been regarded as the wave function equation of universe.
The mini-superspace for this model is a two-dimensional space with
coordinates $(b,\phi )$ with $0<b<\infty $ and $-\infty <\phi
<\infty $. To solve the WD equation, we can follow the separation
of variable method by using the
following assumption%
\begin{equation}
\Psi (b,\phi )=\Phi (\phi )B(b).  \label{Psi}
\end{equation}%
Taking Eq. (\ref{Psi}) into account, it is easy to find that the
WD equation becomes
\begin{equation}
\dfrac{1}{B}\dfrac{d^{2}B}{db^{2}}-\dfrac{12}{\Phi b^{2}}\dfrac{d^{2}\Phi }{%
d\phi ^{2}}+\nu b^{4}-\varrho b^{2}=0,  \label{eq41}
\end{equation}%
where Eq. (\ref{eqsuperpot}) helps us to obtain%
\begin{align}
\nu =& \left[ 96a^{4}(3a^{^{\prime }2}+2aa^{^{\prime \prime
}})+24a^{6}\left( \frac{1}{2}\phi ^{^{\prime }2}+V(\phi )\right)
\right] ,
\label{eqnu} \\
\varrho =& 144a^{4}\kappa .  \label{eqrho}
\end{align}

In order to investigate the WD equation, we have to know the
properties of the superpotential $U(b,w)$ and the corresponding
wave function $\Psi (b,\phi )$. Considering Eq. (\ref{eq41}), one
finds that a regular solution for the wave function in the limit
$b\ll 1$ leads to the fact that the wave function should be $\phi
$-independent since the coefficient of $\partial _{\phi }^{2}$ in
the WD equation diverges. Therefore, the WD equation
reduces to%
\begin{equation}
\dfrac{d^{2}B}{db^{2}}-\varrho b^{2}B=0,
\end{equation}%
with the following exact solution%
\begin{equation}
B(b\ll 1)=c_{1}\sqrt{b}J_{1/4}\left( \dfrac{\sqrt{-\varrho
}b^{2}}{2}\right) +c_{2}\sqrt{b}Y_{1/4}\left(
\dfrac{\sqrt{-\varrho }b^{2}}{2}\right) ,
\end{equation}%
where $c_{1}$ and $c_{2}$ are two integration constants, and $J$
and $Y$ are
two Bessel functions. It is straightforward to find that we have to choose $%
c_{1}=0$ and $c_{2}=-1$ through imposing the tunneling condition%
\begin{equation}
\lim_{b\rightarrow 0}c_{2}\sqrt{b}Y_{1/4}\left( \dfrac{\sqrt{-\varrho }b^{2}%
}{2}\right) =-\dfrac{2c_{2}}{(-\varrho )^{\frac{1}{8}}\Gamma \left( \frac{3}{%
4}\right) }+\mathcal{O}(b),
\end{equation}%
where the wave function goes to the maximum values for vanishing
$b$. In
addition, for the Hartle-Hawking no boundary, we should choose $%
c_{1}=1,c_{2}=0$ since
\begin{equation}
\lim_{b\rightarrow 0}c_{2}\sqrt{b}J_{1/4}\left( \dfrac{\sqrt{-\varrho }b^{2}%
}{2}\right) =\dfrac{c_{1}2^{\frac{3}{2}}3^{\frac{1}{4}}\Gamma \left( \frac{3%
}{4}\right) }{\pi }b+\mathcal{O}(b^{5}),
\end{equation}%
and when $b$ goes to zero the wave function vanishes.

In the limit $b\gg 1$, we can neglect the derivative with respect
to the scalar field $\phi $. Thus, $\phi $ plays the role of a
parameter in the WD equation, and the problem is reduced to the
one-dimensional mini-superspace
model. So, Eq. (\ref{eq41}) reduces to%
\begin{equation}
\dfrac{d^{2}B}{db^{2}}+(\nu b^{4}-\varrho b^{2})B=0,  \label{eqBb}
\end{equation}%
with a semi-classical solution known as the Vilenkin tunneling wave function%
\begin{equation}
B(b)\approx \dfrac{c\exp {\left( \dfrac{i(\nu b^{2}-\varrho )^{(3/2)}}{3\nu }%
\right) }}{b(\nu b^{2}-\varrho )^{(1/4)}},  \label{Vil}
\end{equation}%
which is an oscillatory solution for $\nu b^{2}-\varrho >0$ with a
damped amplitude while for $\nu b^{2}-\varrho <0$ it is only a
damping solution as
\begin{equation}
B(b)\approx \dfrac{c\exp {\left( \dfrac{(\varrho -\nu b^{2})^{(3/2)}}{3\nu }%
\right) }}{b(\varrho -\nu b^{2})^{(1/4)}}.  \label{damp}
\end{equation}%
These are the analogue of a wave-packet peaked about a classical
particle trajectory in the ordinary quantum mechanics.\newline The
exact solution of Eq. (\ref{eqBb}) in the region of $\nu /\varrho
b^{2}>1 $ where superpotential is positive is
\begin{eqnarray}
B(b)&=&c_{3}e^{f(b,\kappa )}HenuT(\lambda ,\delta ,\eta ,\zeta b)+\nonumber  \\
&&c_{4}e^{-f(b,\kappa )}HenuT(\lambda ,\delta ,\eta
,-\zeta b),
\end{eqnarray}%
where%
\begin{align*}
f(b,\kappa )=& \dfrac{(-0.33)(-1.5\varrho +\nu b^{2})b}{\sqrt{-\nu }}, \\
\lambda & =\dfrac{-(0.16+0.28I)\varrho ^{2}}{(-\nu )^{(\frac{4}{3})}}, \\
\delta & =0, \\
\eta & =\dfrac{(-0.57+0.99I)\varrho }{(-\nu )^{\frac{2}{3}}}, \\
\zeta =& (0.44+0.76I)(-\nu )^{\frac{1}{6}}b.
\end{align*}%
It is notable that by converting $\nu \rightarrow -\nu $, $\eta
\rightarrow
-\eta $ and $f\rightarrow -f$, one can obtain the solution for the case of $%
\nu /\varrho b^{2}<1$, where superpotential is positive. 
Besides for $b\gg 1$ the superpotential consists of two terms, a
curvature term $\varrho b^{2}$ and $\nu b^{4}-$term as
\begin{equation}
U(b)=\nu b^{4}-\varrho b^{2}.  \label{eq52}
\end{equation}%
To have a quantum tunneling, the superpotential may have a
maximum. So, we
should have a barrier with the following conditions%
\begin{equation}
\dfrac{dU}{db}=0,\hspace{0.5cm}\dfrac{d^{2}U}{db^{2}}<0.
\end{equation}%
The first condition leads to%
\begin{equation}
\dfrac{dU}{db}=0\hspace{0.5cm}\Rightarrow \hspace{0.5cm}b=b_{bar}=\sqrt{%
\dfrac{\varrho }{2\nu }}
\end{equation}%
in which by inserting it into the second condition, we find
\begin{equation}
\dfrac{d^{2}U}{db^{2}}\mid _{b=b_{bar}}<0\hspace{0.5cm}\Rightarrow \hspace{%
0.5cm}4\varrho <0,
\end{equation}%
and the height of barrier is%
\begin{equation}
U(b=b_{bar})=U_{max}=-\dfrac{\varrho ^{2}}{4\nu }.
\end{equation}%
In order to have a positive $U_{max}$, we should consider a limitation as $%
\nu <0$. Finally, the conditions for having a quantum tunneling are%
\begin{equation}
\nu <0\text{ \ \ and \ \ }\varrho <0.
\end{equation}%
The condition $\varrho <0$ imposes $\kappa <0$ while for the
condition $\nu <0$, we should look at $\nu $ in more details since
it depends on the kind of thick brane. One may also look at the
tunneling rate between the turning points $b=0$ and $b=\sqrt{\nu
/\varrho }$ which is approximately given by
\begin{eqnarray}
|T|^{2}&\approx& \exp {\left( -2\int_{0}^{\sqrt{\nu /\varrho
}}\sqrt{\nu
b^{4}-\varrho b^{2}}db\right) } \nonumber \\
&\approx& \exp {\left( \dfrac{2(-\varrho )^{3/2}%
}{3\nu }\left[ 1-\left( \dfrac{\nu ^{2}}{\varrho ^{2}}-1\right) ^{(3/2)}%
\right] \right) }.
\end{eqnarray}%
In the special cases, we get%
\begin{align}
|T|^{2}& \approx \exp {\left( \dfrac{2\nu ^{2}}{3(-\varrho )^{3/2}}\right) },%
\hspace{0.5cm} & \nu ^{2}/\varrho ^{2}& \gg 1 \\
|T|^{2}& \approx \exp {\left( -8\sqrt{-\kappa }\right)
},\hspace{0.5cm} &
\nu ^{2}/\varrho ^{2}& =1 \\
|T|^{2}& \approx \exp {\left( \dfrac{2(-\varrho )^{(3/2)}\left( 1-i\right) }{%
3\nu }\right) },\hspace{0.5cm} & \nu ^{2}/\varrho ^{2}& \ll 1
\end{align}%
It is obvious that for the first two cases, the rate of tunneling
is real but it is imaginary for the third case, and therefore, the
condition $\nu ^{2}/\varrho ^{2}\ll 1$ does not occur.\newline
Now, we will more investigate quantum tunneling by considering
thick branes with two different four-dimensional solutions in the
case of $\phi (w,t)=\phi (w)$.

\subsection{Model 1: Four-dimensional de-Sitter thick brane solution}

The stable de-Sitter thick brane solution with a scalar field
potential is given by
\cite{Wang:2002pka},\cite{Dzhunushaliev:2009va}
\begin{equation}
V(\phi )=V_{0}\cos ^{2(1-\sigma )}\left( \dfrac{\phi }{\phi _{0}}\right) ,%
\hspace{0.5cm}\phi _{0}=\sqrt{3\sigma (1-\sigma )},
\end{equation}%
where $V_{0}$ and $\sigma $ are arbitrary constants. The $Z_{2}$%
-symmetric solution is also calculated as%
\begin{equation}
a=\dfrac{1}{\left( \cosh \left( \dfrac{Hw}{\sigma }\right) \right) ^{\sigma }%
},\hspace{0.5cm}\phi =\phi _{0}\arcsin \left( \tanh \left(
\dfrac{Hw}{\sigma }\right) \right) ,
\end{equation}%
where $H^{2}=\frac{2\sigma V_{0}}{3(1+3\sigma )}$ and $0<\sigma
<1$. In the limit $\sigma \rightarrow 0$, the solution approaches
a de-Sitter thin brane embedded in a five-dimensional Minkowski
bulk.\newline By using Eqs. (\ref{eqnu}) and (\ref{eqrho}) in the
limit of $w=0$, one can
obtain%
\begin{equation}
\nu =\dfrac{12V_{0}(3\sigma -11)}{3\sigma
+1},\hspace{0.5cm}\varrho =144\kappa ,
\end{equation}%
where we have to consider the constraint $0<\sigma <1$ to have
quantum
tunneling. Accordingly, the effective potential is%
\begin{equation}
U=144\kappa b^{2}\left({\Theta}-1\right) ,
\end{equation}%
where ${\Theta}=\dfrac{V_{0}(3\sigma -11)b^{2}}{12\kappa \left(
1+3\sigma \right) }$, and the barrier is characterized as%
\begin{eqnarray*}
b_{bar} &=&\frac{2.45b}{2}\sqrt{\dfrac{1}{3{\Theta}}}, \\
U_{max} &=&\dfrac{-(3555.69\kappa +10667.08\kappa \sigma
+9504V_{0}-2592\sigma V_{0})b^{4}}{144\kappa \left( 1+3\sigma
\right) {\Theta}^{2}}.
\end{eqnarray*}%
As a result, the wave functions are%
\begin{equation}
\begin{array}{cc}
\begin{array}{c}
\psi \approx \dfrac{c_{1}\exp {\left( \dfrac{108ib^{2}\kappa ^{\frac{1}{2}%
}\left( {\Theta}-1\right) ^{\frac{3}{2}}}{25{\Theta}}%
\right) }}{b\left[ 144\kappa \left( {\Theta}-1\right) \right]
^{(1/4)}}, \\
\end{array}
& {\Theta}>1 \\
\begin{array}{c}
\psi \approx \dfrac{c_{1}\exp {\left( \dfrac{108b^{2}\kappa ^{\frac{1}{2}%
}\left( 1-{\Theta}\right) ^{\frac{3}{2}}}{25{\Theta}}%
\right) }}{b\left[ 144\kappa \left( 1-{\Theta}\right) \right]
^{(1/4)}}, \\
\end{array}
& {\Theta}<1%
\end{array}%
.
\end{equation}%
After some manipulations, we can find that the tunneling rate is
\begin{widetext}
\begin{equation}
|T|^{2}\approx \exp \left( {\dfrac{-0.055b^{2}\left( (-144\kappa )^{(3/2)}-%
\left[ 144\kappa \left( \dfrac{{\Theta}^{2}}{b^{4}}-1\right)
\right] ^{(3/2)}\right) }{12\kappa {\Theta}}}\right) .
\end{equation}
\end{widetext}
in which for the case of ${\Theta}^{2}\gg b^{4}$, we have%
\begin{equation}
|T|^{2}\approx \exp \left( \dfrac{8V_{0}^{2}(3\sigma
-11)^{2}}{(-\kappa )^{(3/2)}}\right) .
\end{equation}

\subsection{Model 2: Four-dimensional Minkowski thick brane solution}

Here, we consider the following superpotential \cite{Dzhunushaliev:2009va}%
\begin{equation}
W(\phi )=c\sin (b\phi ).
\end{equation}%
Taking such a superpotential into account, one can find the following $Z_{2}$%
-symmetric and stable solution for the sine-Gordon scalar
potential, warp
function and scalar field%
\begin{align}
V(\phi )=& \dfrac{3}{2}c^{2}\left[ 3b^{2}\cos ^{2}(b\phi )-4\sin ^{2}(b\phi )%
\right] , \\
a(w)=& \left[ \cosh (cb^{2}w)\right] ^{\frac{-1}{3b^{2}}}, \\
\phi (w)=& \dfrac{2}{b}\arctan \left[ \tanh \left( \dfrac{3cb^{2}w}{2}%
\right) \right] .
\end{align}%
Notably, in the limit $c\rightarrow 0$, the
solution approaches a Minkowski thin brane embedded in a
five-dimensional AdS bulk with an effective cosmological constant
$\Lambda =-6c^{2}$. By
using Eqs. (\ref{eqnu}) and (\ref{eqrho}) for $w=0$, we find%
\begin{equation}
\nu =88c^{2}b^{2},\hspace{0.5cm}\varrho =144\kappa .
\end{equation}%
According to what is obtained above, it is obvious that there is
no quantum tunneling since $\nu $ is positive.

\section{The Cosmology of Thick Brane\label{Cos}}

In the following, to have a deep insight for the case of $\kappa
<0$, we
study the cosmology in brane. So, by using of the ansatz metric (\ref%
{eqmetric}), the explicit form of Einstein field equation
(\ref{eqmotion1}) and the equation of motion for the scalar field
(\ref{eqmotion2}) which are arisen form the variation of action
can be written as are \cite{Ahmed:2013lea}
\begin{widetext}
\begin{equation}
\begin{array}{cc}
\dfrac{3}{a^{2}}\left( a^{^{\prime }2}+aa^{^{\prime \prime }}-\dfrac{\dot{b}%
^{2}}{b^{2}}-\dfrac{\kappa }{b^{2}}\right) =-\dfrac{1}{2a^{2}}\dot{\phi}^{2}-%
\dfrac{1}{2}\phi ^{^{\prime }2}-V, & tt-component \\
\dfrac{1}{a^{2}}\left( -2\dfrac{\ddot{b}}{b}-\dfrac{\dot{b}^{2}}{b^{2}}-%
\dfrac{\kappa }{b^{2}}+3a^{^{\prime }2}+3aa^{^{\prime \prime }}\right) =%
\dfrac{1}{2a^{2}}\dot{\phi}^{2}-\dfrac{1}{2}\phi ^{^{\prime }2}-V,
&
ij-component \\
\dfrac{3}{a^{2}}\left( -\dfrac{\ddot{b}}{b}+2a^{^{\prime }2}-\dfrac{\dot{b}%
^{2}}{b^{2}}-\dfrac{\kappa }{b^{2}}\right) =\dfrac{1}{2a^{2}}\dot{\phi}^{2}+%
\dfrac{1}{2}\phi ^{^{\prime }2}-V, & ww-component \\
\ddot{\phi}+3\dfrac{\dot{b}}{b}\dot{\phi}+a^{2}\dfrac{\partial
V}{\partial \phi }-a^{2}\phi ^{^{\prime \prime }}-4a^{^{\prime
}}a\phi ^{^{\prime }}=0.
& scalar\text{ }field\text{ }equation%
\end{array}
\label{Eq40}
\end{equation}
\end{widetext}

Here, we consider two possibility for $\phi $. As the first one,
we look at the case of $\phi (w,t)=\phi (w)$ and then, we consider
$\phi (w,t)=\phi (t)$ as the second case.

\subsection{Case I: $\protect\phi (w,t)=\protect\phi (w):$}

Considering that the scalar field and the warp function depend
only on the extra dimension, one can simplify the relations that
are combined in Eq. (\ref{Eq40}) with the following forms
\begin{widetext}
\begin{align}
tt-component:\hspace{0.5cm}& \dfrac{\dot{b}^{2}}{b^{2}}+\dfrac{\kappa }{b^{2}%
}=\dfrac{a^{2}}{3}\left[ 3\left( \dfrac{a^{^{\prime }2}}{a^{2}}+\dfrac{%
a^{^{\prime \prime }}}{a}\right) +\dfrac{1}{2}\phi ^{^{\prime
}2}+V\right] ,
\label{eq44} \\
ij-component:\hspace{0.5cm}& 2\dfrac{\ddot{b}}{b}+\dfrac{\dot{b}^{2}}{b^{2}}+%
\dfrac{\kappa }{b^{2}}=a^{2}\left[ 3\left( \dfrac{a^{^{\prime }2}}{a^{2}}+%
\dfrac{a^{^{\prime \prime }}}{a}\right) +\dfrac{1}{2}\phi ^{^{\prime }2}+V%
\right] ,  \label{eq55} \\
ww-component:\hspace{0.5cm}& \dfrac{\ddot{b}}{b}+\dfrac{\dot{b}^{2}}{b^{2}}+%
\dfrac{\kappa }{b^{2}}=\dfrac{a^{2}}{3}\left[ 6\dfrac{a^{^{\prime }2}}{a^{2}}%
-\dfrac{1}{2}\phi ^{^{\prime }2}+V\right] ,  \label{eq66} \\
scalar\text{ }field\text{ }equation:\hspace{0.5cm}& \phi
^{^{\prime \prime }}+4\dfrac{a^{^{\prime }}}{a}\phi ^{^{\prime
}}-\dfrac{\partial V}{\partial \phi }=0,  \label{eq77}
\end{align}
\end{widetext}
where the left (right) hand side depends only on $t$ ($w$). So we
can obtain the following set of equations for $b(t)$
\begin{align}
\dfrac{\dot{b}^{2}}{b^{2}}+\dfrac{\kappa }{b^{2}}=& C_{t},  \label{eqfre1} \\
\dfrac{2\ddot{b}}{b}+\dfrac{\dot{b}^{2}}{b^{2}}+\dfrac{\kappa
}{b^{2}}=&
C_{x},  \label{eqfre2} \\
\dfrac{\ddot{b}}{b}+\dfrac{\dot{b}^{2}}{b^{2}}+\dfrac{\kappa
}{b^{2}}=& C_{w},  \label{eqfre3}
\end{align}%
where $C_{t,x,w}$ are unknown constants. The consistency condition
for these equations guides us to find a relation between such
unknown constants
\begin{equation}
C_{w}=\dfrac{C_{x}+C_{t}}{2}.
\end{equation}%
On the other hand, from the right hand side of these equations,
one can obtain
\begin{equation}
C_{x}=3C_{t},
\end{equation}%
and consequently, we can rewrite all the constants in term of
$C_{w}$ as
\begin{equation}
C_{t}=\dfrac{1}{2}C_{w}=\dfrac{1}{2}\lambda ,\hspace{1cm}C_{x}=\dfrac{3}{2}%
C_{w}=\dfrac{3}{2}\lambda  \label{eqlambda}
\end{equation}%
where $\lambda =C_{w}$ is a constant. Now, combination of Eqs. (\ref{eqfre1}%
)-(\ref{eqfre3}) leads to the following differential equations%
\begin{equation}
\dfrac{\ddot{b}}{b}-\dfrac{\dot{b}^{2}}{b^{2}}-\dfrac{\kappa
}{b^{2}}=0, \label{eqfree}
\end{equation}%
\begin{equation}
\dfrac{\dot{b}^{2}}{b^{2}}+\dfrac{\kappa }{b^{2}}-\dfrac{\lambda
}{2}=0. \label{eqfre}
\end{equation}%
By solving Eq. (\ref{eqfree}), one can find%
\begin{equation}
b(t)=\dfrac{c_{1}}{2}\left( e^{\dfrac{t+c_{2}}{c_{1}}}+\kappa e^{-\dfrac{%
t+c_{2}}{c_{1}}}\right) .  \label{eqscal}
\end{equation}%
Inserting the scale factor (\ref{eqscal}) into the conditional equation (\ref%
{eqfre}), we find%
\begin{equation}
c_{1}=\sqrt{\dfrac{2}{\lambda }},
\end{equation}%
and therefore, the scale factor can be rewritten as%
\begin{equation}
b(t)=\sqrt{\dfrac{2}{\lambda }}\left( \dfrac{e^{\sqrt{\frac{\lambda }{2}}%
(t+c_{2})}+\kappa e^{-\sqrt{\frac{\lambda
}{2}}(t+c_{2})}}{2}\right) ,
\end{equation}%
where $c_{2}$ is an integration constant and $\kappa =-1$. It is
worth mentioning that in the case of $\kappa =-1$ and $\lambda
<0$, the scale factor becomes
\begin{equation*}
b(t)=\sqrt{\dfrac{2}{\lambda }}\sin \left( \sqrt{\dfrac{2}{\lambda }}%
(t+c_{2})\right)
\end{equation*}%
which corresponds to the cyclic universe. Also, for $\kappa =-1$
with positive $\lambda $, we find
\begin{equation*}
b(t)=\sqrt{\dfrac{2}{\lambda }}\sinh \left( \sqrt{\dfrac{2}{\lambda }}%
(t+c_{2})\right) .
\end{equation*}%
%

\begin{figure}[tbp]
\hspace{0.4cm} \centering
\subfigure{\includegraphics[width=0.7\columnwidth]{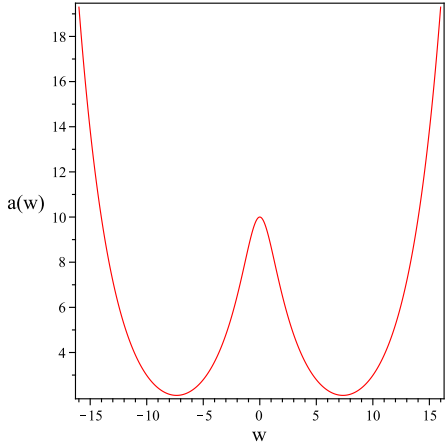}}
\caption{Numerical plot of warp function in terms of $w$ for $\protect%
\lambda =-1$. } \label{figwarp}
\end{figure}

Now, we want to consider the $w$-dependent part of solutions that are%
\begin{align}
\dfrac{a^{2}}{3}\left[ 3\left( \dfrac{a^{^{\prime }2}}{a^{2}}+\dfrac{%
a^{^{\prime \prime }}}{a}\right) +\dfrac{1}{2}\phi ^{^{\prime
}2}+V\right]
=& C_{t}, \\
a^{2}\left[ 3\left( \dfrac{a^{^{\prime
}2}}{a^{2}}+\dfrac{a^{^{\prime \prime
}}}{a}\right) +\dfrac{1}{2}\phi ^{^{\prime }2}+V\right] =& C_{x}, \\
\dfrac{a^{2}}{3}\left[ 6\dfrac{a^{^{\prime
}2}}{a^{2}}-\dfrac{1}{2}\phi ^{^{\prime }2}+V\right] =& C_{w},
\end{align}%
where by considering Eq. (\ref{eqlambda}), we can rewrite them as%
\begin{align}
3\left( \dfrac{a^{^{\prime \prime }}}{a}+\dfrac{a^{^{\prime }2}}{a^{2}}%
\right) -\dfrac{3\lambda }{2a^{2}}=& -\dfrac{1}{2}\phi ^{^{\prime
}2}-V,
\label{eqq68} \\
\dfrac{6a^{^{\prime }2}}{a^{2}}-\dfrac{3\lambda }{a^{2}}=&
\dfrac{1}{2}\phi
^{^{\prime }2}-V,  \label{eqq69} \\
\phi ^{^{\prime \prime }}+4\dfrac{a^{^{\prime }}}{a}\phi ^{^{\prime }}-%
\dfrac{\partial V}{\partial \phi }=& 0.  \label{eqq70}
\end{align}%
Now, by using Eqs. (\ref{eqq68}) and (\ref{eqq69}) and some
algebraic calculations, one can obtain
\begin{align}
V=& -\dfrac{3}{2}\left( \dfrac{a^{^{\prime \prime
}}}{a}+\dfrac{3a^{^{\prime
}2}}{a^{2}}-\dfrac{3\lambda }{2a^{2}}\right) ,  \label{eqqv} \\
\phi ^{^{\prime }2}=& 3\left( \dfrac{a^{^{\prime }2}}{a^{2}}-\dfrac{%
a^{^{\prime \prime }}}{a}-\dfrac{\lambda }{2a^{2}}\right) .
\label{eqqph}
\end{align}%
It is obvious that obtaining an exact analytical solution for the
above equations is hard. Therefore, at first we try to obtain a
numerical solution for the warp functions by assuming the
following ansatz for kink like scalar field
\begin{equation}
\phi =\tanh (w).  \label{eqqp}
\end{equation}%
Inserting Eq. (\ref{eqqp}) into Eq. (\ref{eqqph}) and using ODE
plot, one can plot Fig. \ref{figwarp} for the warp function in the
case of negative $\lambda $.
Now, we try to obtain an analytical approximation solution for the
large and small values of $w$. The corresponding approximate
equations are
\begin{align}
& 3\dfrac{a^{^{\prime \prime }}}{a}+\phi _{s}^{^{\prime }}=0,\hspace{1cm}%
w\rightarrow 0  \label{eqaprox} \\
& \dfrac{3\lambda }{2a^{2}}+\phi _{l}^{^{\prime }}=0,\hspace{1cm}%
w\rightarrow \infty .
\end{align}%
and
\begin{align}
& \dfrac{3a^{^{\prime \prime }}}{2a}+V_{s}=0,\hspace{1cm}w\rightarrow 0 \\
& \dfrac{9\lambda }{4a^{2}}-V_{l}=0,\hspace{1cm}w\rightarrow
\infty .
\end{align}

\begin{figure*}[tbp]
\hspace{0.4cm} \centering
\subfigure[]{\includegraphics[width=0.8\columnwidth]{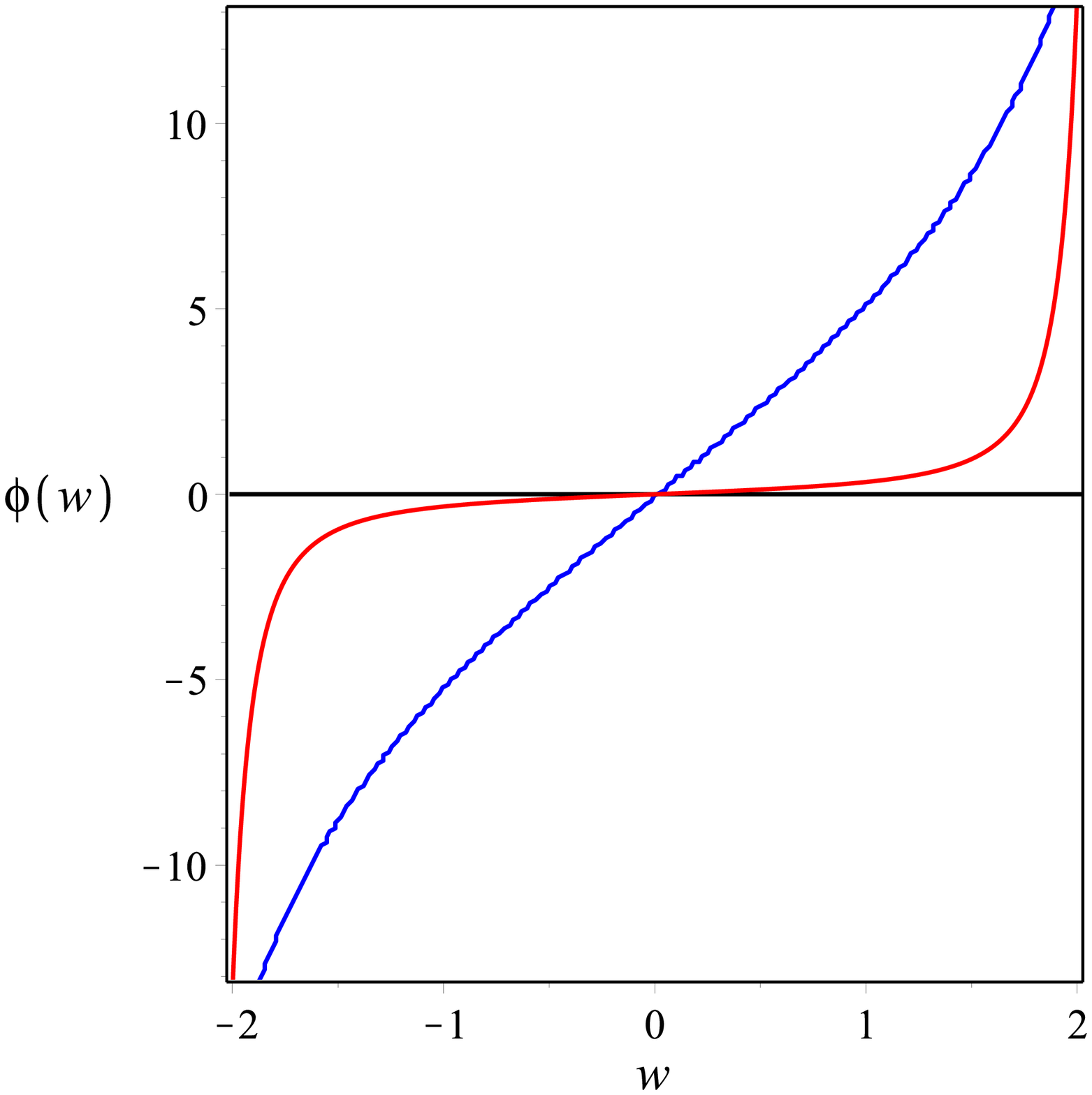}} %
\subfigure[]{\includegraphics[width=0.8\columnwidth]{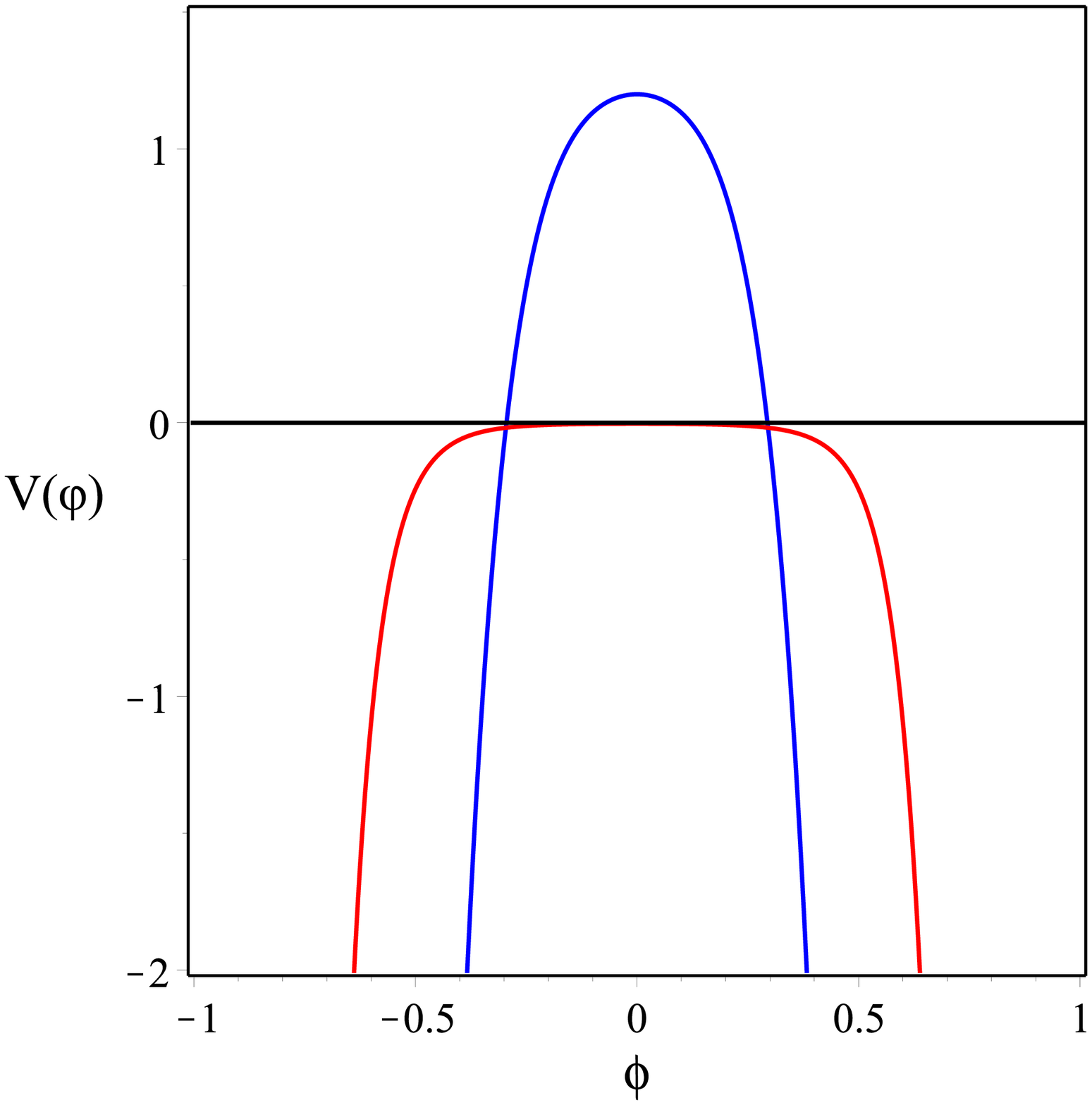}}
\caption{Plot of $\protect\phi _{s}$ (blue line) and $\protect\phi
_{l}$ (red line) in terms of $w$ for $\protect\lambda =-1$ and
$\protect\alpha =5$ (left) plot of $V_{s}$(blue line) and $V_{l}$
(red line) in terms of $w$ for $\protect\lambda =-1,\protect\alpha
=5$ (right).} \label{figbt}
\end{figure*}

In order to solve these equations according to Fig. \ref{figwarp} for $%
\lambda =-1$, we assume the following ansatz for the warp function%
\begin{equation}
a(w)=(w^{2}-\alpha ^{2})^{2}.  \label{waan}
\end{equation}%
By inserting (\ref{waan}) into Eq. (\ref{eqaprox}), one can get%
\begin{eqnarray}
\phi _{s}&=& \dfrac{24}{{\alpha }}\arctan \left( \dfrac{w}{{\alpha
}}\right),   \\
\phi _{l}&=& \dfrac{\lambda }{\alpha ^{6}(w^{2}-\alpha
^{2})^{3}}\left( \alpha w(33\alpha ^{4}-40\alpha
^{2}w^{2}+15w^{4})\right. \nonumber \\
&&\left. -15(w^{2}-\alpha
^{2})^{3}\arctan (\dfrac{w}{\alpha })\right) .
\end{eqnarray}
and
\begin{align}
V_{s}=& \dfrac{6}{\alpha ^{2}}\left[ 3-2\cosh ^{2}(\alpha \phi
)\right]
\cosh ^{2}(\alpha \phi ), \\
V_{l}=& \dfrac{9\lambda }{4\alpha ^{8}}\cosh ^{8}(\alpha \phi ).
\end{align}%
In figures (\ref{figbt}), we have plotted the behavior of above
functions. As can be seen in the figure (\ref{figbt}a), the
approximate scalar field has mirror symmetry $\phi(w)=-\phi(-w)$.
The scalar potential plotted in the Fig.  (\ref{figbt}a). It can
be seen that the potential is an even function of $w$ ($V (w) = V
(-w)$), but it is unbounded from below. This is a consequence of choosing the warp function \eqref{waan} as $w^4$ when $\vert w\vert\to \infty$.

\subsection{Case II: $\protect\phi (w,t)=\protect\phi(t):$}

By considering a time dependent scalar field, it is
straightforward to simplify the relations of Eq. (\ref{Eq40}) as
follow
\begin{widetext}
\begin{align}
tt-component:\hspace{0.5cm}& \dfrac{3}{a^{2}}\left( a^{^{\prime
}2}+aa^{^{\prime \prime }}-\dfrac{\dot{b}^{2}}{b^{2}}-\dfrac{\kappa }{b^{2}}%
\right) =-\dfrac{1}{2a^{2}}\dot{\phi}^{2}-V,  \label{eq56} \\
ij-component:\hspace{0.5cm}& \dfrac{1}{a^{2}}\left( -2\dfrac{\ddot{b}}{b}-%
\dfrac{\dot{b}^{2}}{b^{2}}-\dfrac{\kappa }{b^{2}}+3a^{^{\prime
}2}+3aa^{^{\prime \prime }}\right)
=\dfrac{1}{2a^{2}}\dot{\phi}^{2}-V,
\label{eq57} \\
ww-component:\hspace{0.5cm}& \dfrac{3}{a^{2}}\left( -\dfrac{\ddot{b}}{b}%
+2a^{^{\prime }2}-\dfrac{\dot{b}^{2}}{b^{2}}-\dfrac{\kappa }{b^{2}}\right) =%
\dfrac{1}{2a^{2}}\dot{\phi}^{2}-V,  \label{eq58} \\
scalar\text{ }field\text{ }equation:\hspace{0.5cm}& \ddot{\phi}+3\dfrac{\dot{%
b}}{b}\dot{\phi}+a^{2}\dfrac{\partial V}{\partial \phi }=0.
\label{eq59}
\end{align}
\end{widetext}
Subtracting Eq. (\ref{eq57}) from Eq. (\ref{eq58}), we can obtain%
\begin{equation}
\dfrac{\ddot{b}}{b}+\dfrac{2\dot{b}^{2}}{b^{2}}+\dfrac{2\kappa }{b^{2}}%
-3a^{^{\prime }2}+3aa^{^{\prime \prime }}=0.  \label{eq60}
\end{equation}%
Separation of time-dependent and $w-$dependent terms, we find%
\begin{align}
\dfrac{\ddot{b}}{b}+\dfrac{2\dot{b}^{2}}{b^{2}}+\dfrac{2\kappa
}{b^{2}}=& c,
\label{eqb61} \\
3a^{^{\prime }2}-3aa^{^{\prime \prime }}=& c,  \label{eqa62}
\end{align}%
where $c$ is the separation constant. Now, we can solve Eq.
(\ref{eqa62}) to obtain
\begin{equation}
a(w)=c_{1}\sqrt{c}\sin \left(
\dfrac{w+c_{2}}{\sqrt{3}c_{1}}\right) . \label{eqa63}
\end{equation}%
in which $c_{1}$ and $c_{2}$ are two integration constants.

Considering Eqs. (\ref{eq56}), (\ref{eq58}) and (\ref{eqb61}), we can find%
\begin{equation}
\dot{\phi}^{2}=6\left( \dfrac{\dot{b}^{2}}{b^{2}}+\dfrac{\kappa }{b^{2}}-%
\dfrac{c}{3}\right) .  \label{eq62}
\end{equation}%
As the final step, we use Eq. (\ref{eq58}), to achieve a constant
scalar
potential as follows%
\begin{equation}
V=-\dfrac{6a^{^{\prime \prime }}}{a}=\dfrac{2}{3c_{1}}.
\end{equation}%
which is supported for the flat region of slow-roll inflation. By
inserting the above potential in Eq. (\ref{eq59}) and applying the
slow-roll inflation criteria ($\dot{\phi}\ll V,\ddot{\phi}\ll
3H\dot{\phi}$), we can find the initial value for the scalar field
$\phi =\phi _{0}$.

\begin{figure*}[tbp]
\hspace{0.4cm} \centering
\subfigure[]{\includegraphics[width=0.8\columnwidth]{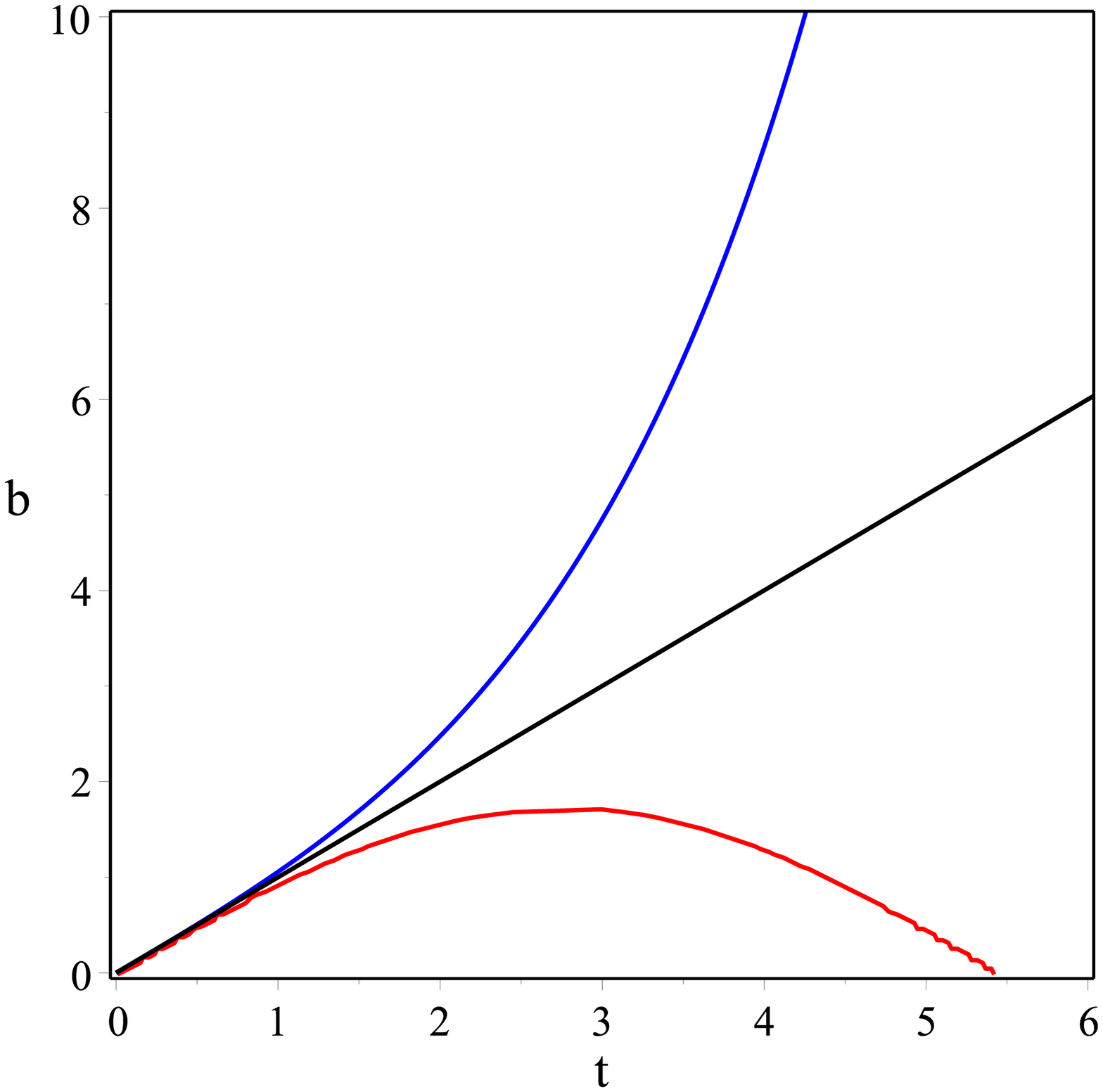}} %
\subfigure[]{\includegraphics[width=0.8\columnwidth]{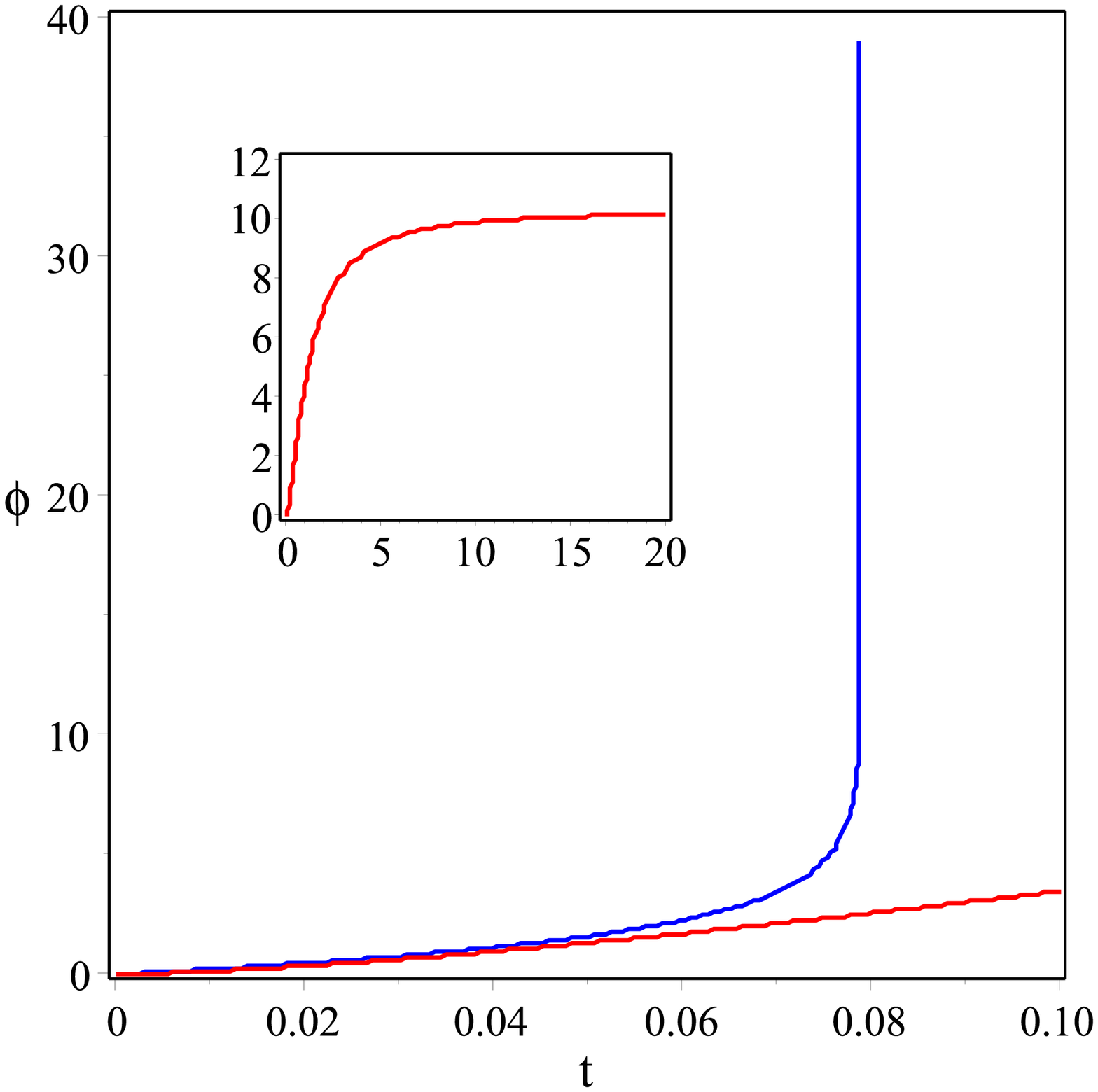}}
\caption{Plots of scale factor in terms of $t$ for $\protect\kappa =-1$ and $%
c=-1$(red line), $c=0$ (black line) and $c=1$ (blue line)(left) plots of $%
\protect\phi $ in terms of $t$ for $\protect\kappa =-1,c=1$ (blue line) and $%
c=-1$ (red line) and in the inset for the large value has been
plotted (right).} \label{figbt2}
\end{figure*}

According to Eqs. (\ref{eqb61}) and (\ref{eq62}), we plot Fig.
\ref{figbt2} to show the behavior of both the scale factor and
scalar field in terms of
time for $\kappa =-1$ and different values of $c$. According to Eq. (\ref%
{eqa63}), only the case of $c=1$ is acceptable since $c=0,-1$
leads to zero and imaginary warp functions. As can be seen in
figure (\ref{figbt2}a), the scale factor in terms of time has an
inflationary behavior ($b\propto t^2$). Also, such a behavior one
can see in figure (\ref{figbt2}b) for scalar field in terms of
time. This behavior is similar to the inflaton field in high
energy \cite{Choudhury:2011sq}.

\section{Conclusion\label{Con}}

In this paper, we have presented five-dimensional thick brane
solutions supported by a scalar field. We have briefly addressed
the formalism of canonical gravity and the WD equation as applied
to the brane. We have seen that only in the case of $ \kappa=-1 $,
tunneling occurs which means that the appropriate classical
cosmology subject to quantization is the spatially paraboloid
case.

Then, we studied the cosmology of thick brane for $\kappa=-1$ and
in the case of time (in)dependent scalar field. In each case, we
have obtained the corresponding solutions of both the scale factor
and scalar field. It is worth mentioning that considering a
time-dependent scalar field leads to a constant scalar potential
which is supported by the known slow-roll inflation.

We also leave the case where $\phi(w,t)=\phi(w)\varphi(t)$ in
appendix for future consideration.

\appendix
\section{Case: $\phi(w,t)=\phi(w)\varphi(t)$}
Here we consider the case $\phi(w,t)=\phi(w)\varphi(t)$.
By using of scalar field equation (\ref{Eq40}), we have
\begin{align}\label{eqA2}
\dfrac{\dot{\varphi}(t)}{\varphi(t)}+\dfrac{3\dot{b}(t)}{b(t)}\dfrac{\dot{\varphi}(t)}{\varphi(t)}-&\left(\dfrac{a^2 \phi^{''}(w)}{\phi(w)}+4a a^{'}\dfrac{\phi^{'}(w)}{\phi(w)}\right)=\nonumber\\
&-\dfrac{a^2}{\varphi(t)\phi(w)}\dfrac{\partial V}{\partial \phi(w,t)}.
\end{align}
Also, by adding $tt$ and $ww$ components of field equations (\ref{Eq40}), one can obtain
\begin{equation}\label{eqA3}
\dfrac{3}{a^2}\left(3a^{'2}+aa^{''}-\dfrac{\ddot{b}}{b}-\dfrac{2\dot{b}^2}{b^2}-\dfrac{2\kappa}{b^2}\right)=-2V.
\end{equation}
In order to solve the field equation (\ref{eqA2}) and
(\ref{eqA3}), we should separate the equations in terms of $t$ and
$w$. To this end, one should consider appropriate choices for
$V(\phi)$. As a simple case we considered $V=0$. In this case the
field equations reduce to
\begin{align}\label{App2}
&\dfrac{\ddot{b}}{b}+\dfrac{2\dot{b}^2}{b^2}+\dfrac{2\kappa}{b^2}-c=0\nonumber\\
&3a^{'2}+aa^{''}-c=0
\end{align}
and
\begin{align}\label{App3}
&\ddot{\varphi}(t)+\dfrac{3\dot{b}}{b}\dot{\varphi}(t)-c^{'}\varphi(t)=0\nonumber\\
&\phi^{''}(w)+\dfrac{4a^{'}}{a}\phi^{'}(w)-\dfrac{c^{'}}{a^{2}}\phi(w)=0.
\end{align}
One can solve the equations (\ref{App2}) for warp function and
scale factor and then insert them in to the equations (\ref{App3})
obtain the scalar fields.

\end{document}